\documentclass[epj]{svjour}
\usepackage{graphics}
\usepackage{amsmath}
\usepackage{mathtools, cuted}
\usepackage{lipsum, color}

\begin{document}
\title{Nuclear excitations within microscopic EDF approaches : pairing and temperature effects on the dipole response}
\titlerunning{Nuclear excitations within microscopic EDF approaches}

\author{E. Y\"{u}ksel \inst{1}, G. Col\`o \inst{2,3}, E. Khan \inst{4}, \and Y. F. Niu \inst{5,6}}
   
\authorrunning{E. Y\"{u}ksel \textit{et. al}.,}
\institute{Department of Physics, Yildiz Technical University, 34220
Esenler, Istanbul, Turkey \and Dipartimento di Fisica, Universit\`a degli 
             Studi di Milano, via Celoria 16, I-20133 Milano, Italy
							\and INFN, Sezione di Milano, Via Celoria 16, 20133 Milano, Italy  \and Institut de Physique Nucl\'eaire, Universit\'e Paris-Sud, IN2P3-CNRS, Universite Paris-Saclay, F-91406 Orsay Cedex, France \and School of Nuclear Science and Technology, Lanzhou University, Lanzhou 730000, China \and
ELI-NP, Horia Hulubei National Institute for Physics and Nuclear Engineering, 30 Reactorului Street, RO-077125,
Bucharest-Magurele, Romania}
\date{Received: date / Revised version: date}
%
\abstract{
In the present work, the isovector dipole responses, both in the resonance region and 
in the low-energy sector, are investigated using the microscopic nuclear Energy Density Functionals (EDFs). The self-consistent QRPA model based on Skyrme Hartree Fock BCS approach is applied to study the evolution of the isovector dipole strength by increasing neutron number and temperature. First, the isovector dipole strength and excitation energies are investigated for the Ni isotopic chain at zero temperature. The evolution of the low-energy dipole strength is studied as a function of the neutron number. In the second part, the temperature dependence of the isovector dipole excitations is studied using the self-consistent finite temperature QRPA, below and above the critical temperatures. It is shown that new excited states become possible due to the thermally occupied states above the Fermi level, and opening of the new excitations channels.
In addition, temperature leads to fragmentation of the low-energy strength around the neutron separation energies, and between 9 and 12 MeV. We find that the cumulative sum of the strength below E$\leq12$ MeV decreases in open-shell nuclei due to the vanishing of the pairing correlations as temperature increases up to T=1 MeV. The analysis of the transition densities in the low-energy region shows that the proton and neutron transition densities display a mixed pattern: both isoscalar and isovector motion of protons and neutrons are obtained inside nuclei, while the neutron transition density is dominant at the surface region.}
%
\maketitle
\section{Introduction}
\label{intro}
Since the discovery of giant dipole resonances in nuclei \cite{bet37,gold48,ste50}, the properties of nuclear excitations have been studied extensively due to their relevance for nuclear structure properties. Although the properties of the giant resonances are well established for nuclei around the stability line, experimental and theoretical works are still needed to investigate their behavior under extreme conditions, i.e., increasing the number of nucleons, the neutron-proton asymmetry, the temperature or the deformation. 
Over the last two decades, the evolution of the isovector dipole resonance with increasing neutron number gained a lot of interest, both experimentally and theoretically. One important result of these studies is the formation of the low-energy strength below the giant dipole resonance region by increasing the neutron or proton number. It is shown that the low-energy dipole strength exhausts a few percent of the isovector dipole strength and is located around the particle threshold \cite{paar00,sav13,bra16}.
Several authors have proposed the name of ``\textit{pygmy dipole resonance}'', especially when one or few states emerge as clear peaks. If the strength is fragmented one should better talk about ``\textit{pygmy dipole strength}'' but in any event a different behavior as compared to the well-known giant dipole resonance is stressed here.
Extensive studies have been performed to elucidate the formation and nature of these low-energy states in nuclei \cite{paar00,sav13,bra16,yuk12,roc12,vre12}. In recent years, the low-energy dipole states have gained particular interest due to their connection with the neutron skin \cite{pie11,car10,bar13,klim07} and symmetry energy \cite{car10,klim07,bar12}. In addition, it has been shown that the radiative neutron capture cross sections and rates are sensitive to the properties of the low-energy dipole states around the neutron separation energies \cite{gori02,lit09}. Therefore, detailed structure and behavior of these low-energy states are quite important for the predictions of the astrophysical processes which take place at finite temperatures.

The giant resonances in highly excited (hot) nuclei have also been the subject of many studies with the aim of investigating the response of nuclei under extreme conditions \cite{bor98}.
Experimentally, they have been studied either using inelastic scattering of light particles or
fusion-evaporation reactions in order to investigate the changes of the width of the giant dipole resonance \cite{bra89,ram96,bau98,kel99,kell99,heck03,wie06,muk12,pan12,mon18}. Since both temperature and spin can impact the width of the giant dipole resonance, the studies have been mainly carried out to disentangle these two effects. 
It has been shown that both the spin and temperature lead to an increase in the width of the giant dipole resonance, and the temperature effects start to become visible above critical temperatures T $> 1.5$ MeV \cite{san06}.

Theoretical works have also been devoted to study temperature effects on nuclear excitations. Among various theoretical models, the quasiparticle random phase approximation (QRPA) is known as an appropriate tool to investigate the excited state properties of nuclei. Although it cannot make predictions for the width of the resonances or detailed structure of the low-energy excitations, it enables us to make systematic and reliable calculations along the nuclear chart. The extensions of the RPA to include the finite temperature effects were made long time ago, and the finite temperature RPA was applied to investigate the nuclear excitations in various multipolarities \cite{civi84,vaut84,bes84,saga84}. In Ref. \cite{yifei09}, the self-consistent calculations were performed using the relativistic DD-ME2 functional and it was shown that the temperature effect leads to an increase of the low-energy strength for monopole and dipole excitations. Using the Skyrme functionals and finite temperature continuum-QRPA in coordinate space, nuclear excitations were also studied in tin and oxygen nuclei \cite{khan04}. Recently, the self-consistent finite temperature QRPA (FT-QRPA) was developed to study the nuclear excitations in open-shell nuclei below the critical temperatures \cite{yuk17}. The first application of the self-consistent FT-QRPA was made to study the dipole and quadrupole excitations in $^{68}$Ni and $^{120}$Sn. It was shown that the low-energy dipole strength is fragmented and the formation of the new excited states become possible in the low-energy region due to the scattering of nucleons to the higher single-particle states at T$>1.0$ MeV. Recently, the finite temperature relativistic time blocking approximation (FT-RTBA) also became available for the calculations of the nuclear excitations \cite{lit18}. 

In this work, the isovector dipole response of Nickel isotopes are studied by increasing neutron number and temperature. The Skyrme-type SLy5 \cite{sly5} and SkI3 \cite{SKI3} energy density functionals are employed in the calculations. The self-consistent finite temperature QRPA is used in the calculations, which also takes into account pairing effects in open-shell nuclei below the critical temperatures. A special emphasis is given to the temperature effect on the low-energy excited states of nuclei, which is relevant for the astrophysical processes. 

The paper is organized as follows. In Sec. \ref{sec:1}, the details of the calculations are given and the finite temperature QRPA is summarized. In Sec. \ref{sec:2}, the results are presented for the isovector dipole excitations in the Ni isotopes, at zero and finite temperature. The effect of the neutron excess and temperature in the low-energy region is discussed by considering the configurations of the excited states, cumulative sums of the strengths and transition densities. The summary and conclusions are given in Sec. \ref{sec:4}.

\section{Formalism}
\label{sec:1}
In the present work, the ground-state properties of nuclei are calculated using the finite temperature Hartree Fock BCS approach (FT-HFBCS) (see Refs. \cite{good81,yuk14} for more information). We use a zero-range density-dependent pairing interaction of surface type \cite{ber91}. The pairing strength $V_{0}$ is taken as 920 (1000) MeV$\cdot$fm$^{3}$ for the SLy5 (SkI3) interaction, which provides reasonable pairing gap values at zero temperature for Nickel nuclei \cite{sat98}. Using the Skyrme-type SLy5 (SkI3) interaction, the neutron pairing gap values are determined as $\Delta_{n}$ =1.36 (1.49), 1.60 (1.87) and 1.83 (1.62) MeV for $^{58}$Ni, $^{62}$Ni and $^{66}$Ni, respectively. At finite temperature, the occupation probabilities of the states are given by
\begin{equation}
n_i=v_{i}^{2}(1-f_{i})+u_{i}^{2}f_{i},
\end{equation} 
where $u_i$ and $v_i$ are the BCS amplitudes. The temperature dependent Fermi-Dirac distribution function is given by
\begin{equation}
\label{eq:fd}
f_{i}=[1+e(E_{i}/k_{B}T)]^{-1},
\end{equation}
where $E_{i}$ is quasiparticle (q.p.) energy, $k_{B}$ is the Boltzmann constant, and T is the temperature \cite{yuk17}. Due to the use of the grand-canonical description in our model, a sharp pairing phase transition is expected from superfluid state to the normal state at critical temperatures \cite{good81,yuk14,niu13}. As we discuss below, the critical temperature values for the pairing phase transition are generally obtained above T$_{c}$$>$0.7 MeV in open-shell Nickel nuclei. However, the nucleus is a finite system and effects due to the particle number fluctuations should be also taken into account. Several works have been devoted to solve this issue in the finite temperature mean-field theory using the particle-number projected BCS \cite{gam13,dan07}, the Lipkin-Nogami method in the static-path approximation \cite{dan93} or the Landau theory \cite{god84,al88}. 
It was shown that the pairing effects still persist at T$\geq$1 MeV, albeit they are small, and the sharp pairing phase transitions are removed. 
We should mention that inclusion of effects due to the particle number fluctuations can be important for the calculations below T$<$1 MeV due to the persistence of the pairing correlations. However, considering the low impact of the pairing correlations that persist at T$\geq$1MeV, our model can be used to explore the changes in the excitation spectrum, at least in a qualitative way. 

At zero temperature, the QRPA can be applied on top of the Hartree Fock BCS calculations to calculate the nuclear excited states of nuclei. The QRPA formalism is well-known and the relevant equations can be found in Refs. \cite{ring80,yuk18}. At finite temperature, proper description of the nuclear excited states requires the extension of the well known QRPA to the FT-QRPA \cite{som83,yuk17}. The FT-QRPA matrix is given by
\begin{equation}
\left( { \begin{array}{cccc}\label{eq:qrpa}
 \widetilde{C} & \widetilde{a} & \widetilde{b} & \widetilde{D} \\
 \widetilde{a}^{+} & \widetilde{A} & \widetilde{B} & \widetilde{b}^{T} \\
-\widetilde{b}^{+} & -\widetilde{B}^{\ast} & -\widetilde{A}^{\ast}& -\widetilde{a}^{T}\\
-\widetilde{D}^{\ast} & -\widetilde{b}^{\ast} & -\widetilde{a}^{\ast} & -\widetilde{C}^{\ast}
 \end{array} } \right)
 \left( {\begin{array}{cc}
\widetilde{P}  \\
\widetilde{X }  \\
\widetilde{Y}  \\
\widetilde{Q} 
 \end{array} } \right)
 = E_{\nu}
  \left( {\begin{array}{cc}
\widetilde{P}  \\
\widetilde{X}  \\
\widetilde{Y}  \\
\widetilde{Q} 
\end{array} } \right), \end{equation}
where $E_{\nu}$ represents the excited state energies. The $\widetilde{P}, \widetilde{X}, \widetilde{Y}$ and $\widetilde{Q}$ are the eigenvectors of the matrix and they are used in the calculation of the strength function. The $A$ and $B$ matrices already exist in the QRPA and contribute both at zero and finite temperature. The other parts of the FT-QRPA matrix start to contribute only by increasing temperature. 

The temperature dependencies of the matrices are given by
\begin{align}
\begin{split}
\widetilde{A}_{abcd}=&(E_{a}+E_{b})\delta_{ac}\delta_{bd} \\
&+\sqrt{1-f_{a}-f_{b}} A'_{abcd}\sqrt{1-f_{c}-f_{d}} \label{eq:temp},
\end{split}\\
\begin{split}
\widetilde{B}_{abcd}=&\sqrt{1-f_{a}-f_{b}} B_{abcd}\sqrt{1-f_{c}-f_{d}}, 
\end{split}\\
\begin{split}
\widetilde{C}_{abcd}=&(E_{a}-E_{b})\delta_{ac}\delta_{bd}\\
&+\sqrt{f_{b}-f_{a}} C'_{abcd}\sqrt{f_{d}-f_{c}},
\end{split}\\
\begin{split}
\widetilde{D}_{abcd}=&\sqrt{f_{b}-f_{a}} D_{abcd}\sqrt{f_{d}-f_{c}}, 
\end{split}\\
\begin{split}
\widetilde{a}_{abcd}=&\sqrt{f_{b}-f_{a}} a_{abcd}\sqrt{1-f_{c}-f_{d}}, 
\end{split}\\
\begin{split}
\widetilde{b}_{abcd}=&\sqrt{f_{b}-f_{a}} b_{abcd}\sqrt{1-f_{c}-f_{d}},
\end{split}\\
\begin{split}
\widetilde{a}_{abcd}^{+}=&\widetilde{a}_{abcd}^{T}=\sqrt{f_{d}-f_{c}} a_{abcd}^{+}\sqrt{1-f_{a}-f_{b}}, 
\end{split}\\
\begin{split}
\widetilde{b}_{abcd}^{T}=&\widetilde{b}_{abcd}^{+}=\sqrt{f_{d}-f_{c}} b_{abcd}^{T}\sqrt{1-f_{a}-f_{b}}.\label{eq:temp1}
\end{split}
\end{align}
In the finite temperature QRPA matrix, the diagonal part of the matrix includes both ($E_{a}+E_{b}$) and ($E_{a}-E_{b}$) configuration energies, and the latter starts to contribute at finite temperature. Detailed information about the structure of the matrices is given in Ref. \cite{yuk17}. After the diagonalization of the FT-QRPA matrix, the reduced transition probability is calculated by

\begin{equation}
\begin{split}
B(EJ)&=\bigl|\langle \nu ||\hat{F}_{J}||\widetilde0\rangle \bigr|^{2}\\
&=\biggl|\sum_{c\geq d}\Big\{(\widetilde{X}_{cd}^{\nu}+ \widetilde{Y}_{cd}^{\nu})(v_{c}u_{d}+u_{c}v_{d})\sqrt{1-f_{c}-f_{d}} \\
&+(\widetilde{P}_{cd}^{\nu}+\widetilde{Q}_{cd}^{\nu})(u_{c}u_{d}-v_{c}v_{d})\sqrt{f_{d}-f_{c}}\Big\}\langle c ||\hat{F}_{J}||d\rangle\biggr|^{2},
\end{split}
\label{bel}
\end{equation}
where $\hat{F}_{J}$ is the operator of the relevant excitation. In addition, $|\nu\rangle$ and $|\widetilde0\rangle$ represent the excited state and the correlated FT-QRPA ground state, respectively. The proton and neutron quasiparticle fractions of an excited state are calculated using the FT-QRPA amplitudes

\begin{equation}
A_{ab}=|\widetilde{X}_{ab}^{\nu}|^{2}-|\widetilde{Y}_{ab}^{\nu}|^{2}+|\widetilde{P}_{ab}^{\nu}|^{2}-|\widetilde{Q}_{ab}^{\nu}|^{2},
\label{Aab}
\end{equation} 
and the normalization condition can be written as
\begin{equation}
\sum_{a\geq b}A_{ab}=1.
\label{aaa}
\end{equation}

The energy weighted moments are given by
\begin{equation}
m_{k}=\sum_{\nu}B(EJ,\widetilde0\rightarrow \nu)E_{\nu}^{k}.
\end{equation}  

In the present study, the dipole excitations in Nickel nuclei are studied using the self-consistent FT-QRPA and the Skyrme-type SLy5 \cite{sly5} and SkI3 \cite{SKI3} interactions. While the SLy5 interaction is optimized to describe the properties of exotic nuclei, the SkI3 interaction is designed to explain the isotopic shift in Pb isotopes. The calculations are performed with the assumption of spherical symmetry and box boundary conditions are used, i.e., the continuum is discretized in a 20 fm box. The quasiparticle energy cut off is taken as E$_{cut}=100$ MeV.
The discrete strength of the dipole response of nuclei is displayed by means of a Lorentzian smoothing, and the Lorentz functions have a width of 1.0 (0.5) MeV at zero (finite) temperature \cite{yuk17}.

\section{RESULTS}
\label{sec:2} 
\subsection{Dipole response in Nickel nuclei at zero temperature}
\label{sec:22} 
As mentioned above, the low-energy strength or pygmy dipole resonance occurs due to the neutron (or proton) excess in nuclei and it is generally located around the particle threshold \cite{paar00,sav13,bra16}. The characteristic behavior of these excited states differ from the well-known giant dipole resonance. Therefore, it has been the subject of many studies over the last two decades \cite{paar00,sav13,bra16,yuk12,roc12}. Experimentally, the low-energy strength was studied in $^{68}$Ni using the virtual photon scattering technique \cite{wie09}. Recently, the low-energy dipole strength was also measured in $^{70}$Ni by using the Coulomb excitation at the RIKEN Radioactive Isotope Beam Factory (RIBF) \cite{wie18}. The total strength and the percentage of the energy weighted
sum rule (EWSR) between 8 and 12 MeV was found as 2.2$\pm$0.4 $e^{2}fm^{2}$ and $4.8\% \pm 0.9\%$, respectively. In addition, comparison of the results showed that the low-energy strength increases with neutron number. While the total strength between 6 and 8 MeV was found as 0.6$\pm$0.22 $e^{2}fm^{2}$ in $^{68}$Ni, this value was estimated as 1.06$\pm$0.14 $e^{2}fm^{2}$ ($1.5\% \pm 0.2\%$ of the EWSR) in $^{70}$Ni \cite{wie18}.

\begin{figure}
\resizebox{0.5\textwidth}{!}{%
  \includegraphics{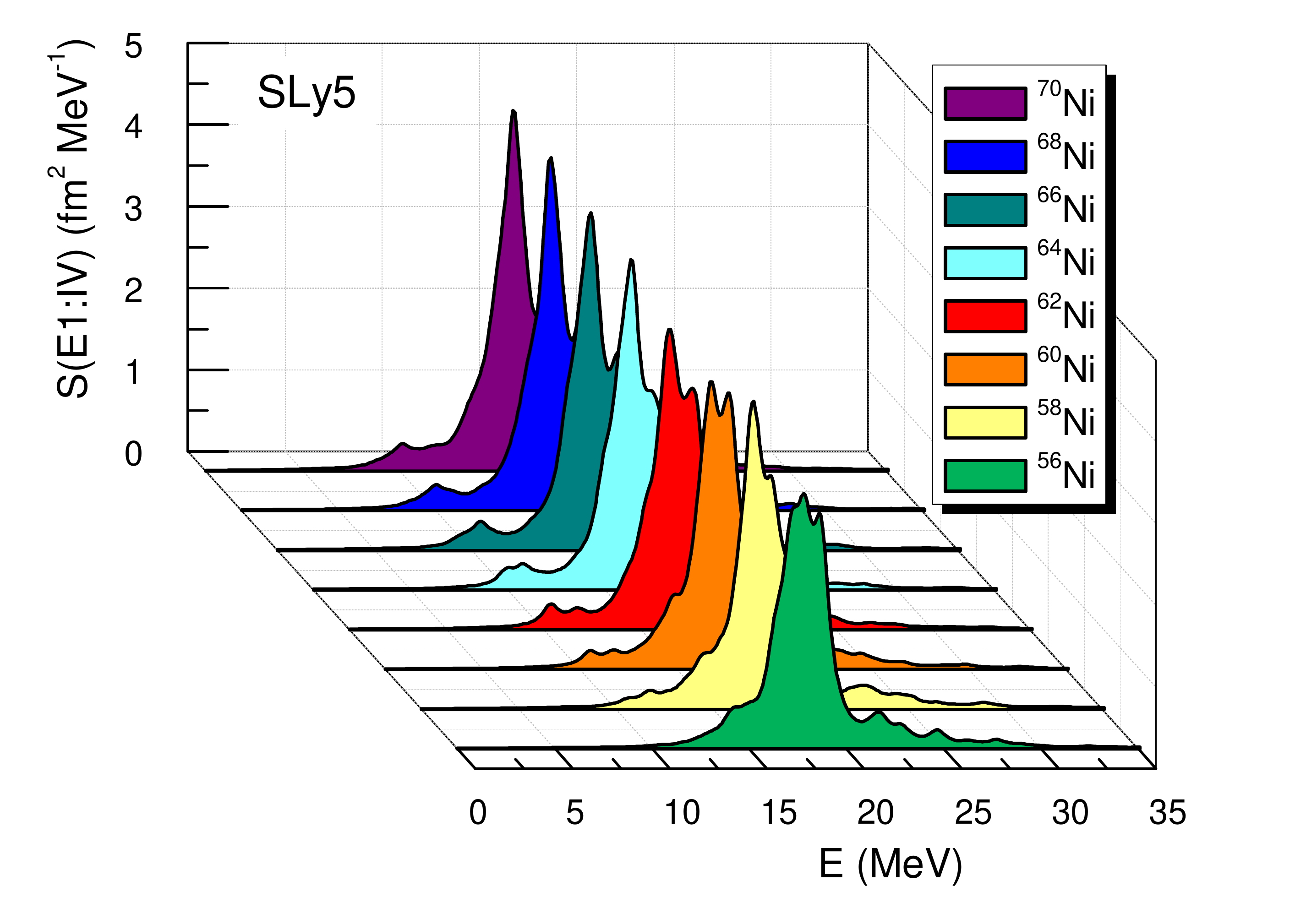}
}
\caption{(Color online) The isovector dipole strength in the Nickel isotopic chain. The calculations are performed with the QRPA using the Skyrme-type SLy5 energy density functional at zero temperature. The discrete isovector dipole states are smoothed with a Lorentzian of $\Gamma=1$ MeV width.}
\label{fig:01}       
\end{figure}

\begin{figure}
\resizebox{0.5\textwidth}{!}{%
  \includegraphics{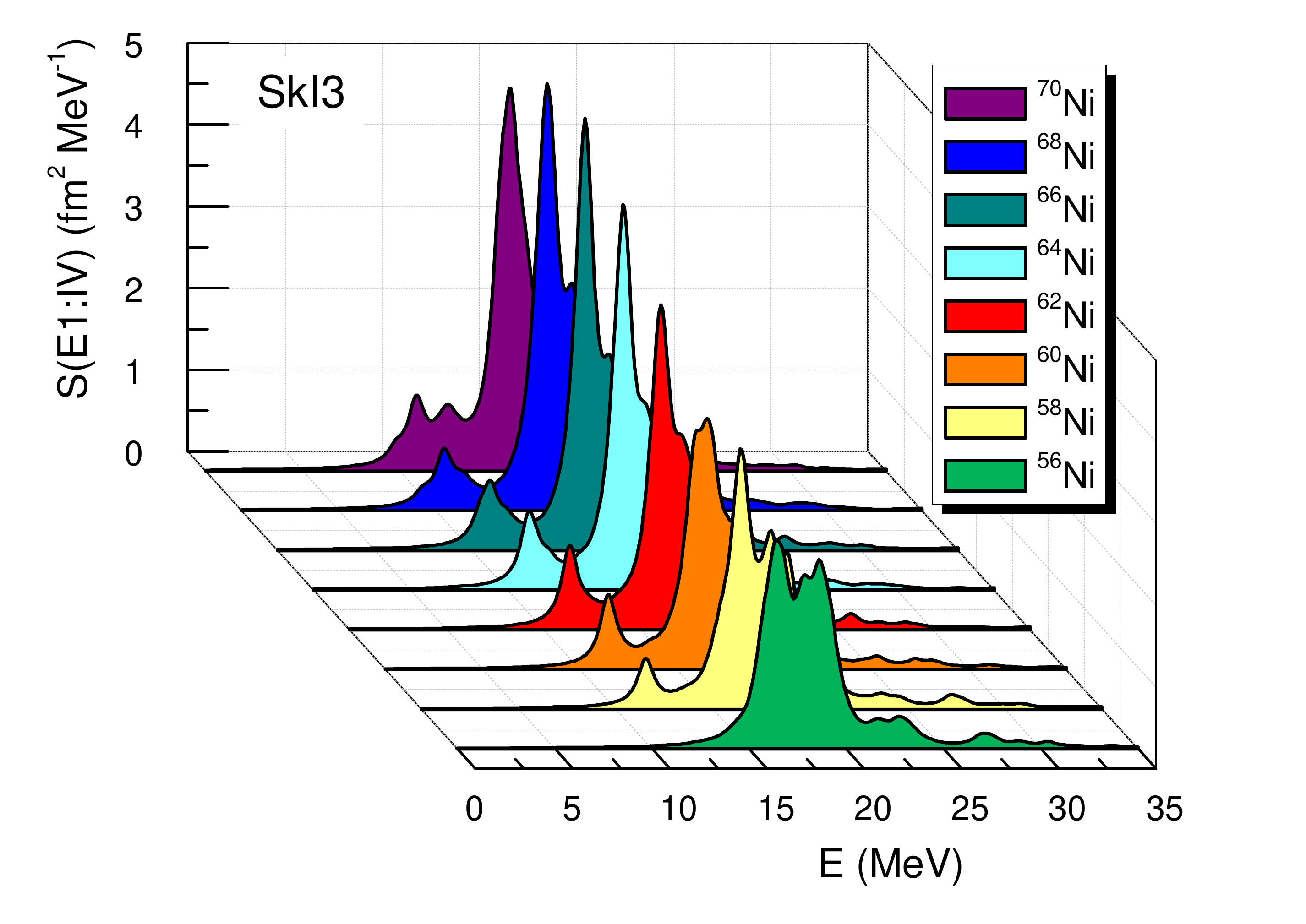}
}
\caption{(Color online) Same as in figure \ref{fig:01}, but using SkI3 interaction.}
\label{fig:0}       
\end{figure}

\begin{table}[ht]
\caption{The major low-energy dipole excitations for $^{70}$Ni nucleus at zero temperature. The two quasiparticle configurations and their contributions to the norm of the states (in percentage) (see Eq. \ref{aaa}) are displayed for each excited state, separately. Here, $\nu$ and $\pi$ represent neutron and proton, respectively.} 
\centering 
\begin{tabular}{c c c c } 
\hline\hline  \\[-1.0em]
  & SkI3 & SLy5\\
\hline \ \\ [-1.ex]
 &E=10.85 MeV & E=10.08 MeV  \\
 Transitions  &B(E1)=0.872 $fm^{2}$  & B(E1)=0.324 $fm^{2}$  \\
\hline \\[-1.0em] 
$\nu2d_{3/2} - \nu2p_{1/2} $  & 35.88  & 1.82\\
$\nu2f_{7/2} - \nu1g_{9/2} $  & 12.34  &  \\
$\nu2d_{5/2} - \nu1f_{5/2} $  & 12.07  & 5.36 \\
$\nu3s_{1/2} - \nu2p_{3/2} $  & 7.31   & \\
$\nu2d_{5/2} - \nu2p_{3/2} $  & 5.18   & 8.31\\
$\nu1g_{9/2} - \nu1f_{7/2}$   & 4.92   & 8.25      \\ 
$\nu2d_{3/2} - \nu1f_{5/2}$   & 4.69   & 60.96\\
$\nu3d_{3/2} - \nu1f_{5/2}$   & 1.23   & 2.11 \\ 
\hline 
$\pi1g_{9/2}\rightarrow \pi1f_{7/2}$ & 5.78 & 4.26  \\ 
\hline \hline \\ [-1.ex]
\end{tabular}
\label{table:xy} 
\end{table}

In figure \ref{fig:01}, the isovector dipole strength is displayed in 
the case of the SLy5 interaction by increasing neutron number. We start 
from $^{56}$Ni, which is doubly magic so that pairing correlations do not play a role, and there we do not obtain any low-energy state below 12 MeV.
By increasing neutron number, the formation of the low-energy strength can be seen in other Nickel nuclei. Using the SLy5 interaction, we do not obtain any excited state below 8 MeV up to $^{70}$Ni. Therefore, comparison of the results with the experimental data is not possible. The absence of the excited states in the low-energy region is related to the limitations of the model used in the calculations. It is well-known that the Q(RPA) method is based on the inclusion of 1p-1h (or two-quasiparticle) configurations, and quite successful in the description of the collective excitations of nuclei in the high-energy region. However, it fails to predict the detailed structure of the excited states in the low-energy region. The proper description of the low-energy states necessitates the inclusion of the higher order configurations (i.e., 2p-2h, 3p-3h, etc.) \cite{gam11} or implementation of the particle-vibration coupling techniques \cite{lit07,roca17,yifei16}. In spite of that, the QRPA can be used to make qualitative predictions on the evaluation of the low-energy states with increasing neutron number.

Considering the states between 8 and 12 MeV, the total strength is obtained as 0.75 $e^{2}fm^{2}$ ($2.77\%$ of the EWSR) in $^{70}$Ni, which is also quite low compared to the recent experimental data. Nonetheless, the low-energy states start to shift downwards while the strength increases by increasing neutron number. The calculations are also performed using the Skyrme-type SkI3 interaction and the results are displayed in figure \ref{fig:0}. We obtain a similar behavior in the low-energy region, namely, we do not obtain any excited state below 8 MeV, and the low-energy strength increases while the excited states start to shift downwards by increasing neutron number. However, the predicted strength in the low-energy region is larger and more concentrated as compared to the results from the SLy5 interaction. In addition, the low-energy strength is clearly separated from the isovector giant dipole resonance region. Using the SkI3 interaction, the total strength between 8 and 12 MeV is obtained as 1.67 $e^{2}fm^{2}$ ($6.33\%$ of the EWSR) in $^{70}$Ni, which is in better agreement with the experimental data.
This difference between the results of the SLy5 and SkI3 functionals can be explained by considering the value of the slope 
of the symmetry energy at saturation point ($L$) \cite{roc12,car10}. 
While the SLy5 interaction has $L=48.27$ MeV, the SkI3 interaction 
is stiffer with $L=100.52$ MeV. It has been shown that the low-energy strength is correlated with the slope of the symmetry energy and functionals with larger values of $L$ predict larger strength in the low-energy region \cite{roc12,vre12,car10}. 

The wave-function of the excited states can also be used to explain the difference in the low-energy strength of two interactions. In table \ref{table:xy}, we analyze the corresponding transitions for the most prominent peaks in $^{70}$Ni. Using the 
SLy5 interaction, the low-energy state at E=10.08 MeV is mainly formed 
with the $\nu2d_{3/2}-\nu1f_{5/2}$ transition. However, using the 
SkI3 interaction, the excited state at E=10.85 MeV displays a more 
collective behavior characterized by the comparable contribution 
of several transitions. 
In addition, the configurations for the low-energy states obtained by using either SLy5 or SkI3 are quite different 
due also to the different shell structure that is predicted.

\subsection{Dipole response in Nickel isotopes at finite temperatures}
\label{sec:3}
In this section, we study the temperature effect on the isovector dipole response of Nickel isotopes. The calculations are performed at T=0, 0.7, 1.0, and 2.0 MeV using the FT-QRPA and Skyrme-type functionals.

\begin{table}[ht]
\caption{The critical temperature (T$_{c}$) values (in MeV) for the selected Nickel isotopes using SkI3 and SLy5 interactions.} 
\centering 
\begin{tabular}{c c c c } 
\hline\hline  \\[-1.0em]
  & SkI3 & SLy5\\
\hline \\[-1.0em] 
 $^{58}$Ni & 0.84  & 0.69\\
 $^{62}$Ni & 1.01  & 0.93 \\
 $^{66}$Ni & 0.82  & 1.0 \\  
\hline \hline \\ [-1.ex]
\end{tabular}
\label{table:zz} 
\end{table}

Before discussing the temperature effect on the isovector dipole strength in nuclei, it is necessary to discuss the changes induced by the temperature on the ground state properties of nuclei. In table \ref{table:zz}, the critical temperature values for the selected Nickel isotopes are presented. The calculations are performed using the FT-HFBCS and Skyrme-type SkI3 and SLy5 interactions.
For open-shell $^{58}$Ni, $^{62}$Ni and $^{66}$Ni nuclei, the critical temperatures are found to be above 0.7 MeV (except for $^{58}$Ni using SLy5 functional). Therefore, pairing correlations still play a role for these nuclei at T=0.7 MeV. At high temperatures, i.e., at T=2.0 MeV, pairing correlations vanish for all open-shell nuclei, and do not contribute to the FT-QRPA matrices. 

\begin{figure}
\resizebox{0.5\textwidth}{!}{%
  \includegraphics{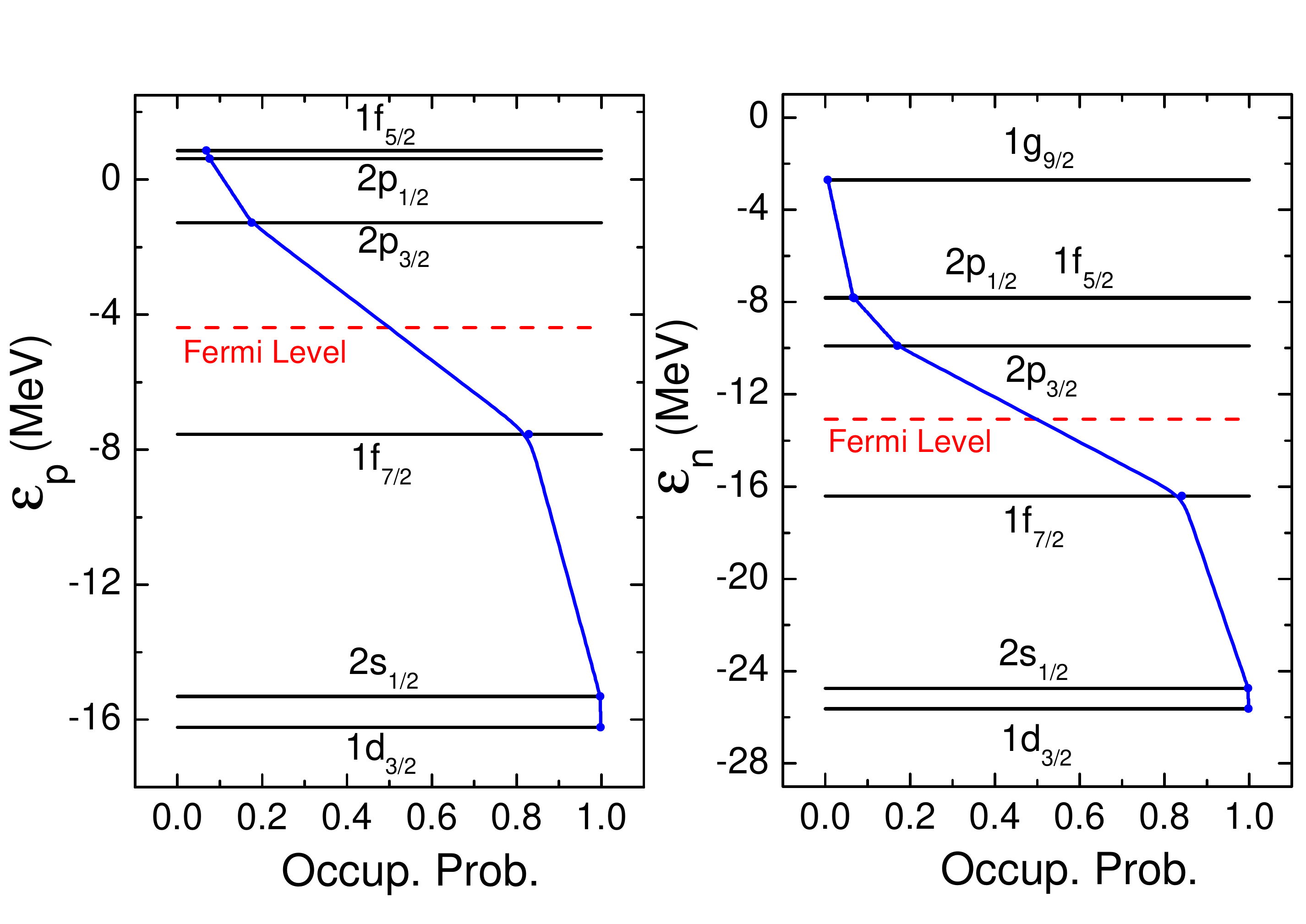}
}
\caption{(Color online) The single-particle energies and occupation probabilities of proton (left panel) and neutron (right panel) states for $^{56}$Ni at T=2 MeV. The calculations are performed using the SkI3 functional.}
\label{fig:spe}       
\end{figure}

By increasing temperature, nucleons are thermally excited to higher energy states. In figure \ref{fig:spe}, the single-particle energies and occupation probabilities of the single-particle states are displayed at T=2 MeV 
using the SkI3 functional. At zero temperature, proton and neutron states are fully occupied up to $1f_{7/2}$ state in doubly-magic $^{56}$Ni nucleus. By increasing the temperature, neutrons are thermally excited from $1d_{3/2}$, $2s_{1/2}$, and $1f_{7/2}$ states below the Fermi level to $2p_{3/2}$, $2p_{1/2}$, $1f_{5/2}$, and $1g_{9/2}$ states above the Fermi level. Similarly, protons are also thermally excited from $1d_{3/2}$, $2s_{1/2}$, and $1f_{7/2}$ states to $2p_{3/2}$ and $2p_{1/2}$ states. A similar picture is obtained in the case of SLy5, despite some quantitative difference due to the lower density of levels (or lower effective mass) associated with SkI3.
At variance with $^{56}$Ni, the neutron $2p_{3/2}$, $2p_{1/2}$, $1f_{5/2}$, and $1g_{9/2}$ states are already partially occupied in open-shell $^{58}$Ni, $^{62}$Ni and $^{66}$Ni nuclei due to the pairing effects at zero temperature. With increasing temperature, the occupation probabilities of the states start to increase (decrease) above (below) the Fermi level. While new excitation channels become possible due to the thermally unblocked states at finite temperature, the excited states are also affected by the changes on the ground state properties (i.e., the q.p energies and occupation probabilities of the proton and neutron states) as well as by the temperature factors in Eqs. (\ref{eq:temp}-\ref{eq:temp1}).

\begin{figure}
\resizebox{0.5\textwidth}{!}{%
  \includegraphics{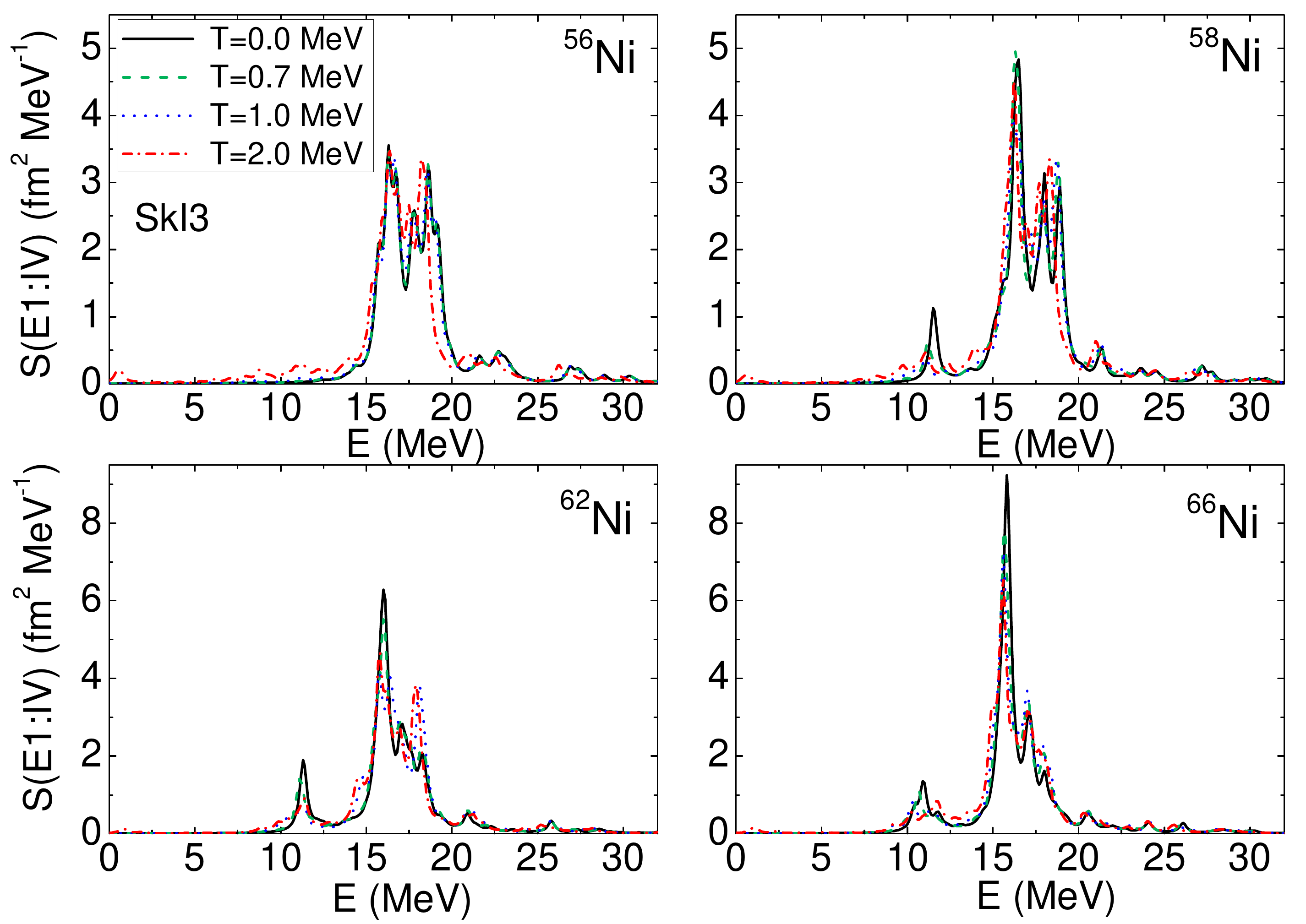}
}
\caption{(Color online) The isovector dipole response in Nickel isotopic chain at T=0, 0.7, 1.0 and 2.0 MeV. The calculations are performed with the FT-QRPA using the Skyrme-type SLy5 energy density functional. The discrete isovector dipole states are smoothed with a Lorentzian of $\Gamma=0.5$ MeV width.}
\label{fig:1}       
\end{figure}

\begin{figure}
\resizebox{0.5\textwidth}{!}{%
  \includegraphics{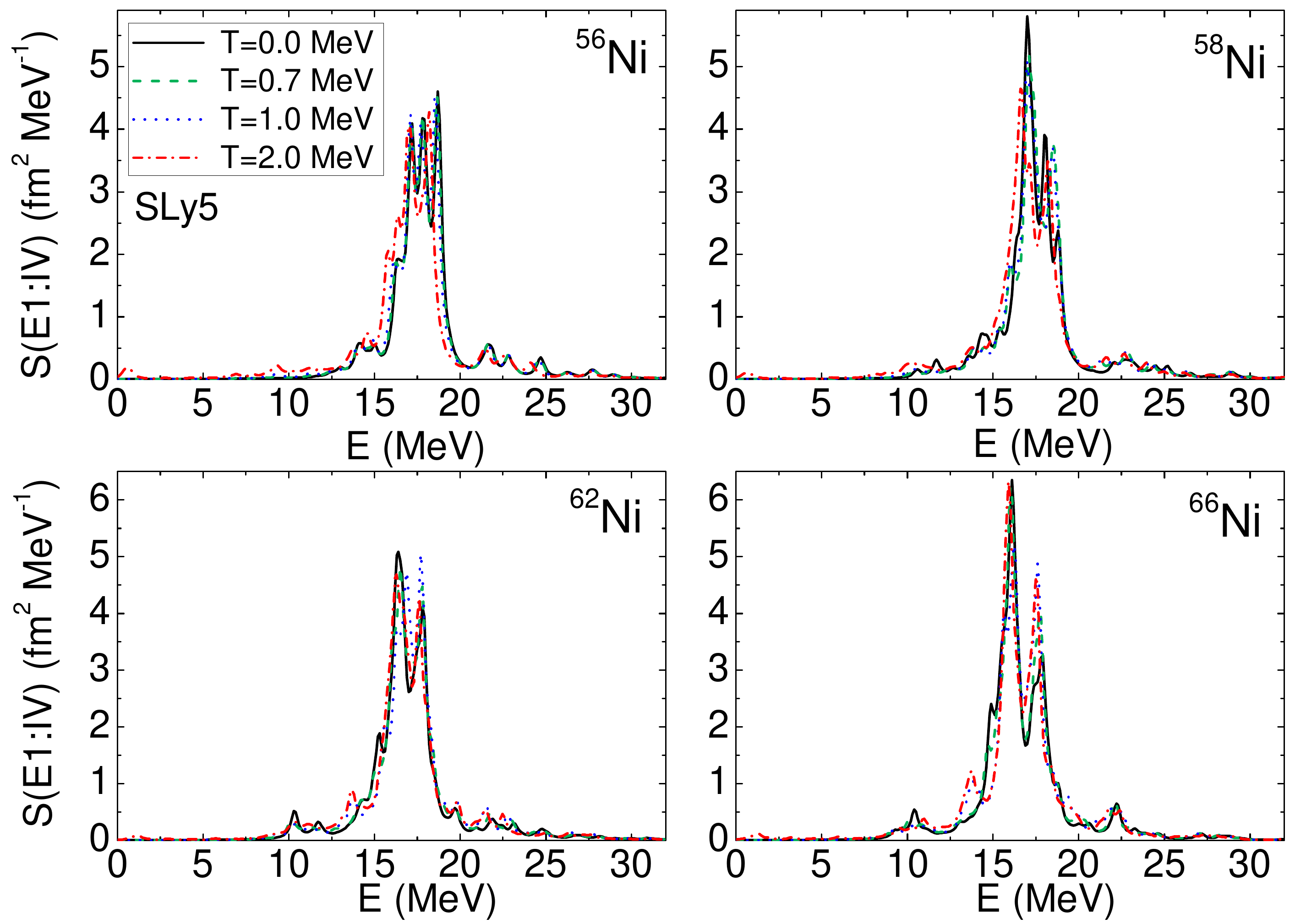}
}
\caption{(Color online) Same as in figure \ref{fig:1} but using SLy5 interaction.}
\label{fig:11}       
\end{figure}

\begin{figure}
\resizebox{0.5\textwidth}{!}{%
  \includegraphics{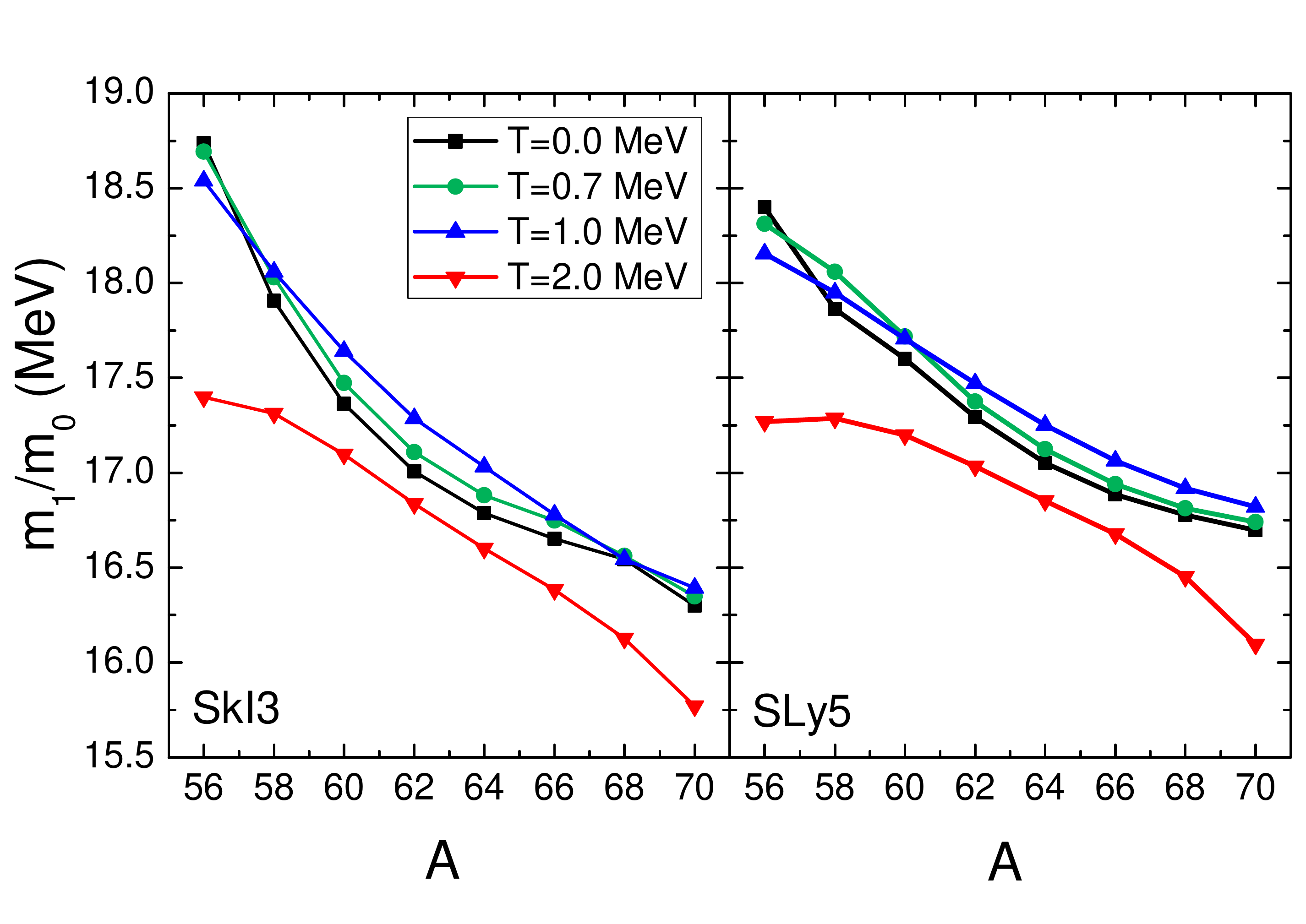}
}
\caption{(Color online) The centroid energy of the isovector dipole response in Nickel isotopes by increasing neutron number at T=0, 0.7, 1 and 2 MeV. The calculations are performed using SkI3 (left panel) and SLy5 (right panel) functionals.}
\label{gdr}       
\end{figure}

In figures \ref{fig:1} and \ref{fig:11}, the temperature effect on the isovector dipole response of $^{56}$Ni,$^{58}$Ni,$^{62}$Ni, and $^{66}$Ni nuclei is illustrated for Skyrme-type SkI3 and SLy5 functionals. The discrete spectra are smoothed with a Lorentzian of $\Gamma=0.5$ MeV width. At T=0.7 MeV, the isovector dipole strength remains almost the same for the selected Nickel isotopes. By increasing temperature further, the isovector dipole states start to shift slightly downwards. In addition, the low-energy states are fragmented and distributed towards lower energies. 
In order to better understand the temperature effect in the isovector dipole spectra, the centroid energies $m_{1}/m_{0}$ are displayed in figure \ref{gdr} as a function of the neutron number at different temperatures. The centroid energies are calculated in the whole excitation energy range, namely by considering the contribution of all the excited states. At zero temperature, the centroid energy is decreasing with increasing neutron number, as expected. Moving to T=0.7 MeV, the centroid energy increases slightly for open-shell nuclei with respect to the case of zero temperature, whereas it decreases slightly in the doubly-magic nucleus $^{56}$Ni. The increase of the centroid energies in open-shell nuclei when temperature increases is related to the weakening of the pairing effects with increasing temperature. Due to the decrease in the strength of the attractive pairing interaction, the centroid energies increase slightly. Above the critical temperature, i.e., T$\geq 1.0$ MeV, pairing effects disappear and the centroid energy continues to increase with temperature in open-shell nuclei, while it decreases in $^{56}$Ni and remains almost constant in $^{68}$Ni. At T=2 MeV, the isovector dipole states continue to shift down and the contribution of the new excitation channels becomes non-negligible in the low-energy region, so that the centroid energy decreases in each nuclei.

\begin{figure}
\resizebox{0.5\textwidth}{!}{%
  \includegraphics{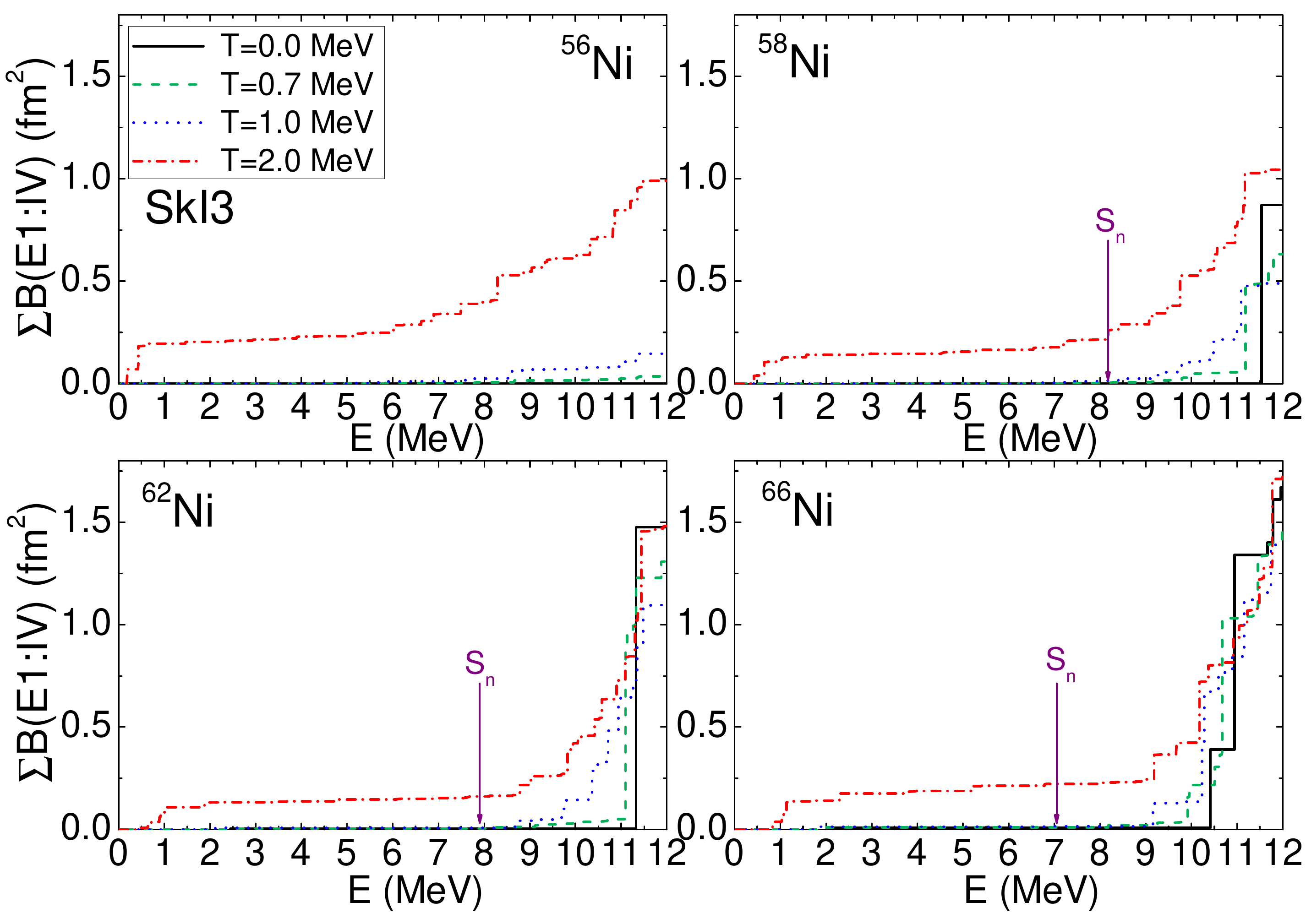}
	}
\caption{(Color online) The cumulative sum of the isovector dipole strength in the low-energy region as a function of the temperature for the $^{56}$Ni,$^{58}$Ni,$^{62}$Ni, and $^{66}$Ni nuclei, using the Skyrme-type SkI3 interaction. The vertical arrows indicate the calculated neutron separation energies.}
\label{fig:2}       
\end{figure}

As mentioned above, the properties of the low-energy states around the separation energy play a major role in the rapid neutron capture processes which take place in stellar environments at finite temperatures. Therefore, investigating the low-energy strength by increasing the temperature is also important for the description of the astrophysical $r$-process.
In figure \ref{fig:2}, the cumulative sum of the isovector dipole strengths is displayed up to E=12.0 MeV using the SkI3 functional. 
Starting with the doubly-magic nucleus $^{56}$Ni at T=0 MeV, we do not obtain any low-energy state below 12.0 MeV. However, by increasing temperature, it is seen that the strength in the low-energy region starts to increase. As explained above, the increase in the low-energy strength is due to the thermal occupation of the states above the Fermi energy, and opening of the new excitation channels at finite temperatures. In table \ref{table:q}, we display the single-particle transitions and their contribution to the wave function of given excited states in the case of $^{56}$Ni at T=2 MeV. It is clear that the newly formed low-energy excitations are made of one or two single neutron (proton) transitions from thermally unblocked states to the continuum. Considering the number of the transitions and their weight, we can say that the low-energy states do not display collectivity.

\begin{table}[ht]
\caption{The selected low-energy excitations for the $^{56}$Ni nucleus at T=2 MeV using SkI3 functional. The excited state energies and transitions are also provided with their contributions (in percentage) to the norm of the states [see Eqs. (\ref{Aab}) and (\ref{aaa})]. Herein, $\pi$ and $\nu$ refer to protons and neutrons, respectively.} 
\centering 
\begin{tabular}{c c c c} 
\hline\hline  \\[-1.0em]
 Excited state energy & Transition & \% \\ [1ex] 
\hline \\[-1.0em] 
0.64 MeV     & $\pi2f_{7/2} - \pi2d_{5/2} $ & 99.76 \\
5.30 MeV     & $\pi3s_{1/2} - \pi2p_{3/2} $ & 99.94 \\ 
6.04 MeV     & $\pi2d_{5/2} - \pi2p_{3/2} $ & 99.16 \\ 
7.96 MeV     & $\pi4s_{1/2} - \pi2p_{3/2} $ & 99.25 \\ 
8.78 MeV     & $\pi3d_{5/2} - \nu2p_{3/2} $ & 96.06 \\ 
             & $\nu2d_{5/2} - \nu1f_{5/2} $ & 2.15 \\ 
9.27 MeV     & $\nu2d_{3/2} - \nu2p_{1/2} $ & 96.71 \\ 
             & $\nu2d_{3/2} - \nu1f_{5/2} $ & 2.51 \\ 	
11.35 MeV    & $\pi5s_{1/2} - \pi2p_{3/2} $ & 53.34 \\ 
             & $\pi4d_{5/2} - \pi2p_{3/2} $ & 6.25 \\ 				
						 & $\nu2d_{3/2} - \nu2p_{3/2} $ & 34.27 \\
\hline\hline \\ [-1.ex]
\end{tabular}
\label{table:q} 
\end{table}

\begin{table*}
\caption{(Color online) The cumulative sum of the isovector dipole strength (in units of $fm^{2}$) in Nickel isotopes below 12.0 MeV. The calculations are performed using the 
SkI3 interaction at T =0, 0.7, 1 and 2 MeV.} 
\centering 
\begin{tabular}{c c c ccc } 
\hline\hline  \\[-1.0em]
  SkI3& T=0.0 MeV & T=0.7 MeV & T=1.0 MeV & T=2.0 MeV     \\
\hline \\[-1.0em] 
 $^{56}$Ni &        & 0.036  & 0.147   & 0.987  \\
 $^{58}$Ni & 0.872  & 0.632  & 0.490   & 1.045  \\
 $^{62}$Ni & 1.475  & 1.308  & 1.097   & 1.480  \\  
 $^{66}$Ni & 1.670  & 1.483  & 1.390   & 1.720  \\  
\hline \hline \\ [-1.ex]
\end{tabular}
\label{table:qq} 
\end{table*}
\begin{table*}[ht]
\caption{Same as in Table \ref{table:qq}, but using the SLy5 interaction.} 
\centering 
\begin{tabular}{c c c ccc } 
\hline\hline  \\[-1.0em]
  SLy5& T=0.0 MeV & T=0.7 MeV & T=1.0 MeV & T=2.0 MeV     \\
\hline \\[-1.0em] 
 $^{56}$Ni & 0.013   & 0.0742  & 0.177   & 0.834  \\
 $^{58}$Ni & 0.301  & 0.230   & 0.310   & 0.778  \\
 $^{62}$Ni & 0.628  & 0.570   & 0.538   & 0.828  \\  
 $^{66}$Ni & 0.670  & 0.641   & 0.626   & 0.935  \\  
\hline \hline \\ [-1.ex]
\end{tabular}
\label{table:q1} 
\end{table*}

As explained in Section \ref{sec:22}, the low-energy strength is already obtained in the open-shell $^{58}$Ni, $^{62}$Ni and $^{66}$Ni nuclei at zero temperature. Below the critical temperature, i.e, at T=0.7 MeV, the low-energy region starts to be fragmented and the state shifts downward slightly. We also find that the cumulative sum of the strength up to E$\leq$12.0 MeV decreases, which is due to the decreasing pairing correlations in open-shell nuclei with increasing temperature. The fragmentation in the low-energy states and the decrease of the cumulative sum of the strength also continue with increasing temperature. At T=1 MeV, the pairing correlations vanish completely, whereas the temperature is not high enough to impact the low-energy strength with the opening of new excitation channels, so that
the cumulative sum of the strength continues to decrease. By further increasing the temperature, at T=2 MeV, the contribution of the new excitation channels becomes non-negligible and the formation of new excited states is obtained below E$\leq$10.0 MeV. In table \ref{table:qq}, we present the cumulative strength below 12.0 MeV in Nickel isotopes. We obtain a considerable dipole strength in the low-energy region of doubly-magic $^{56}$Ni nucleus. For open-shell nuclei, the low-energy strength decreases up to T=1 MeV, and it starts to increase again at higher temperatures with the contribution of the new excitations.  

\begin{figure}
\resizebox{0.5\textwidth}{!}{%
  \includegraphics{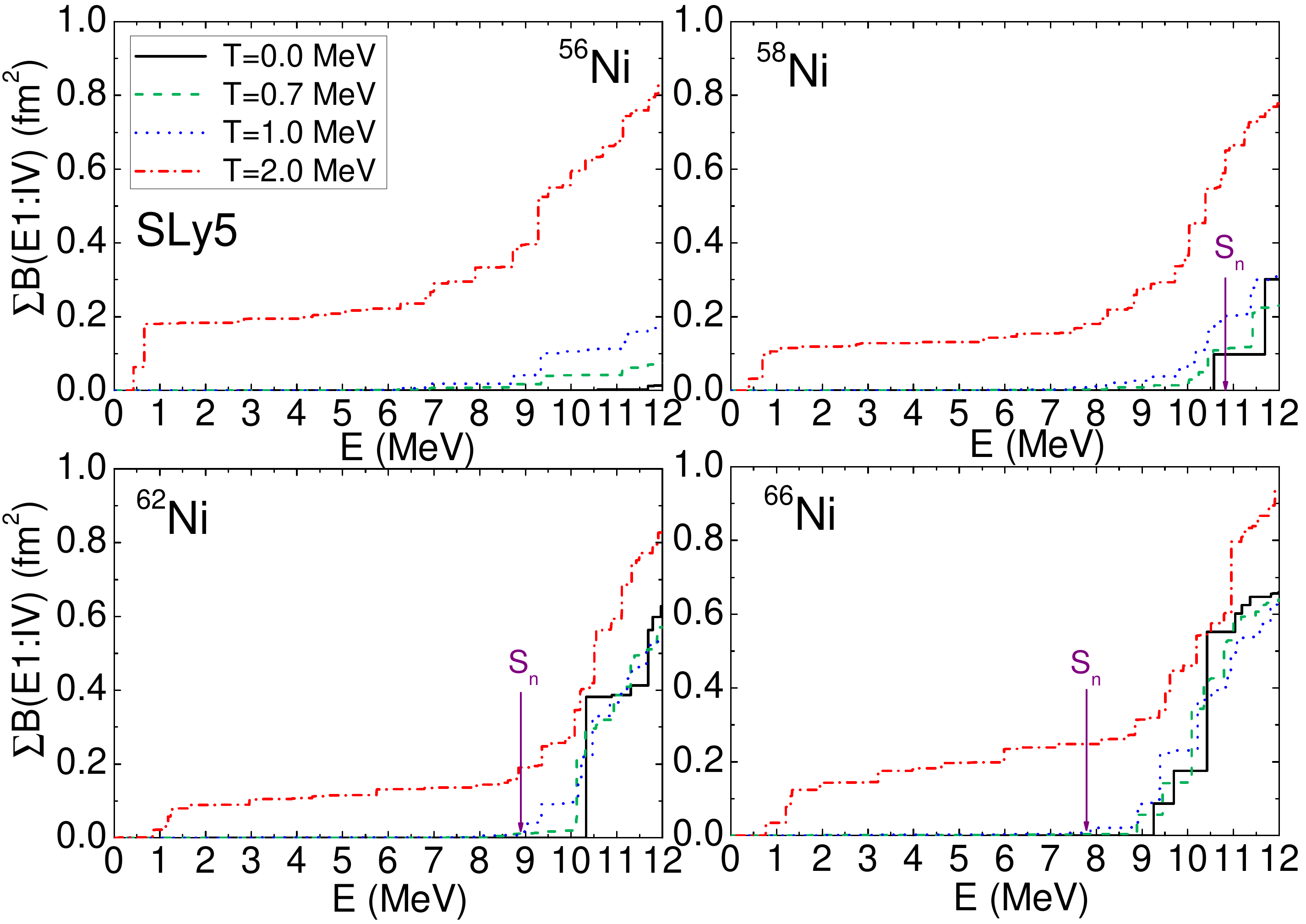}
}
\caption{(Color online) Same as in figure \ref{fig:2} but using the Skyrme-type SLy5 interaction.}
\label{fig:22}       
\end{figure}

In order to test the functional dependence of the results, a similar analysis is also performed using the SLy5 interaction and the results are displayed in figure \ref{fig:22} and table \ref{table:q1}. The changes in the low-energy region with increasing temperature are quite similar as in the case of the SkI3 interaction. In $^{56}$Ni, formation of the low-energy states and increase in the cumulative sum of the strength can be seen by increasing temperature. For open-shell nuclei, the fragmented low-energy states and their integrated strength decrease slightly at T=0.7 MeV. At T=1 MeV, the fragmentation of the low-energy states increase and formation of the new low-energy states are obtained below E$\leq$10.0 MeV. Using the SkI3 functional, the total integrated strength remains almost the same in open-shell nuclei at zero temperature and T=2 MeV. However, the SLy5 functional predicts larger values for the integrated strength at T=2 MeV, as compared to the zero temperature case. This difference can be traced back to the difference in the predicted shell structure. In agreement with the results from the SkI3 functional, the low-energy states are mainly formed with one or two transitions and do not display collectivity when calculated using the SLy5 functional.
\begin{figure*}
\centering
\resizebox{\textwidth}{!}{%
  \includegraphics{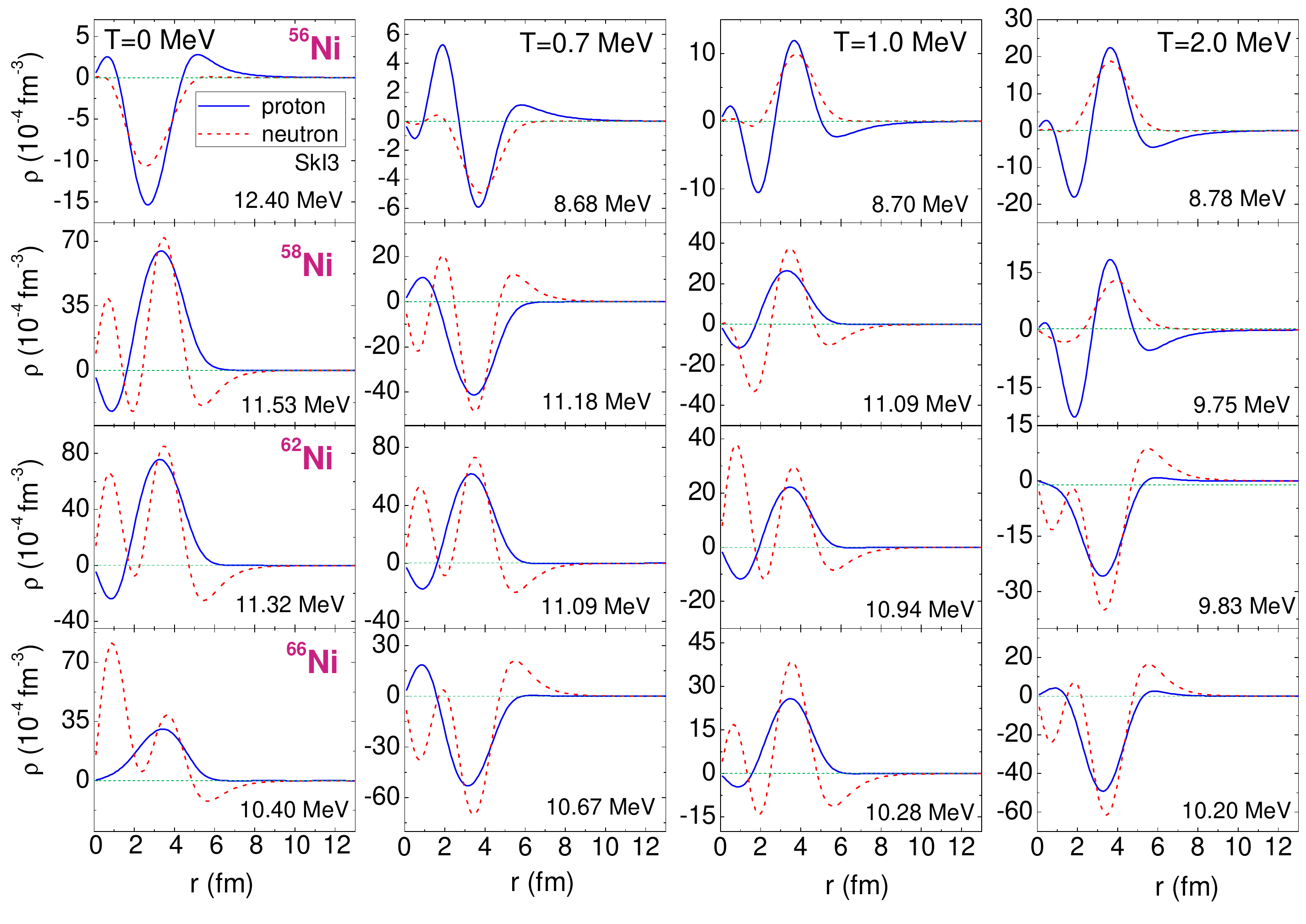}
}
\caption{(Color online) The proton and neutron transition densities associated with the most prominent peaks in the low-energy region of the isovector dipole response at zero and finite temperatures for the nuclei $^{56}$Ni,$^{58}$Ni,$^{62}$Ni, and $^{66}$Ni.}
\label{fig:3}       
\end{figure*}

\subsection{Analysis of proton and neutron transition densities}
The transition densities can also provide useful information about the nature of the low-energy states at zero and finite temperature. The transition density of an excited state is defined using the FT-QRPA amplitudes
\begin{equation}
\begin{split}
	\rho_{\nu}(r)=&\frac{1}{2J+1}\sum_{cd}\Big\{(\widetilde{X}_{cd}^{\nu}+ \widetilde{Y}_{cd}^{\nu})(v_{c}u_{d}+u_{c}v_{d})\sqrt{1-f_{c}-f_{d}} \\
&+(\widetilde{P}_{cd}^{\nu}+\widetilde{Q}_{cd}^{\nu})(u_{c}u_{d}-v_{c}v_{d})\sqrt{f_{d}-f_{c}}\Big\} \\
&\times\langle c ||Y_{J}||d\rangle \frac{u_{c}(r)u_{d}(r)}{r^{2}}
\end{split}
\end{equation}
where $Y_{J}$ are the spherical harmonics and $u$(r) are the radial wave functions. For the analysis of the transitions densities, the summation can be done for protons or neutrons separately. 

The giant dipole resonance region characterised by isovector motion (protons and neutrons oscillate in opposite phase), while the behavior of the low-energy region is more complex, and depends on the chosen interaction. In figure \ref{fig:3}, we display proton and neutron transition densities in $^{56}$Ni, $^{58}$Ni, $^{62}$Ni, and $^{66}$Ni calculated by using the SkI3 functional, at zero and finite temperatures. For presentation purposes, we choose the most prominent peaks in the low-energy region. In $^{56}$Ni, which has equal number of protons and neutrons, the excited states mainly display an isoscalar behavior (in-phase motion of protons and neutrons), with the dominance of protons, both at zero and finite temperatures. By increasing the neutron number and temperature, the excited states display a mixed pattern: both isoscalar and isovector motion can be seen inside the nucleus while neutron transition densities dominate in the surface region. This fast change of the behavior of the transition densities with respect to the neutron number and temperature are a signature of the complex mechanisms at work in the low-energy region of the isovector dipole response.

\section{Conclusion}
\label{sec:4}
In this work, the self-consistent finite-temperature QRPA calculations are performed on top of Hartree-Fock BCS, at zero and finite temperatures. The Skyrme-type energy density functionals SLy5 and SkI3 are used in the calculations. 

First, the isovector dipole response in the Nickel isotopic chain is investigated at zero temperature. It is shown that the QRPA model fails to reproduce the low-energy states below 8 MeV in all nickel nuclei, and comparison of our results with the experimental data is not possible in $^{70}$Ni nucleus due to limitations of the QRPA method in the description of the low-energy states. It is clear that the inclusion of the higher order configurations (or particle-vibration techniques) must be taken into account for the proper description of the states in the low-energy region. However, this issue is presently out of the scope of the present paper.
In spite of its drawbacks, the QRPA predicts the formation of the low-energy strength between 8-12 MeV with increasing neutron number, using both SLy5 and SkI3 functionals. Using the QRPA, the low-energy strength is formed with the contribution of several two q.p. configurations around the Fermi level. Compared to SLy5, the SkI3 functional predicts a larger amount of strength in the low-energy region, which is also well separated from the isovector giant dipole resonance of nuclei. This difference between two functionals is related with the difference in the values of the slope $L$ of the symmetry energy around saturation point. The SkI3 functional has a larger $L$ value and predicts larger strength, which is also consistent with the findings of previous studies \cite{car10,roc12}.

Secondly, the effect of the temperature on the isovector dipole response of Nickel isotopes is investigated. The calculations are performed below (T=0.7 MeV) and above the critical temperatures (T=1 and 2 MeV). While the effect of the temperature is mild up to the T=1 MeV, it is seen that the isovector dipole states start to shift down at higher temperatures. In addition, we find that the centroid energies increase in open-shell nuclei up to T=1 MeV, which is related to the vanishing of the pairing correlations by increasing temperature. By increasing the temperature further, the isovector dipole states shift downward due to the increasing impact of the temperature factors in the FT-QRPA matrices and new low-energy excitations at higher temperatures. 
The effect of the temperature is more pronounced in the low-energy region of the doubly-magic $^{56}$Ni nucleus due to the opening of the new excitation channels at finite temperature. In the open-shell Nickel nuclei, the low-energy dipole states start to be fragmented below the critical temperatures, i.e., at T=0.7 MeV. The formation of new excited states is also obtained below E$\leq$10 MeV by increasing temperature. It is seen that the cumulative sum of the strength decreases in the low-energy region (E$\leq$12 MeV) and below the critical temperatures (T$\leq$1 MeV), using both the SkI3 and SLy5 functionals. The decrease in the cumulative sum of the strength is also related with the vanishing of the pairing correlations with temperature. 

The transition densities of the low-energy dipole states are also analyzed at zero and finite temperatures in the case of the $^{56}$Ni, $^{58}$Ni, $^{62}$Ni, and $^{66}$Ni nuclei. In the doubly-magic $^{56}$Ni nucleus, the new excited states at finite temperatures mainly display a proton dominated isoscalar motion of nucleons. In open-shell nuclei, the low-energy states display both isoscalar and isovector motion inside the nucleus while neutron transition density is dominant in the surface region, both at zero and finite temperatures.

\section*{Acknowledgement}
We are glad to contribute to a volume devoted to the memory of Pier Francesco Bortignon. 
Temperature effects in nuclei have been one of his main interests during his career as a nuclear theorist. He
developed a deep understanding of this topic, from its foundations that are rooted in statistical mechanics
and many-body theory as reported in Ref. \cite{bor98}, to the peculiarities of the phenomena in finite systems like nuclei. He has 
always been strongly supportive of the experimental activities aiming at understanding the limiting temperature for the
nuclear existence, the giant resonances in hot nuclei, the temperature and angular momentum dependence of
nuclear dissipation. Although these topics are not treated in the present contribution, we are proud to 
dedicate our complementary study of pairing vs. temperature effects in low-lying dipole states to his memory,
being certain that if he were alive he would study and criticize it with his usual passion for science.

G.C. acknowledges funding from the European Union’s Horizon 2020 research
and innovation program under Grant No. 654002.

\end{document}